\documentclass{PoS}

\leftline{\parbox{20cm}{\large BI-TP 2010/36, HU-EP-10/60, LU-ITP 2010/007}
\vspace{1cm}}

\title{NSPT study of the three-loop lattice gluon propagator in Landau gauge}

\ShortTitle{NSPT study of the 3-loop gluon propagator in Landau gauge}

\author{\speaker{C. Torrero}\\
        Institut f\"ur Theoretische Physik, Universit\"at Regensburg, D-93040 Regensburg, Germany\\
        Dipartimento di Fisica, Universit\`a di Pisa, I-56127 Pisa, Italy\\
        E-mail: \email{christian.torrero@df.unipi.it}}

\author{F. Di Renzo\\
        Dipartimento di Fisica, Universit\`a di Parma \& INFN, I-43100 Parma, Italy\\
        E-mail: \email{francesco.direnzo@fis.unipr.it}}

\author{E.-M. Ilgenfritz\\
        Fakult\"at f\"ur Physik, Universit\"at Bielefeld, D-33501 Bielefeld, Germany\\ 
        Institut f\"ur Physik, Humboldt-Universit\"at zu Berlin, D-12489 Berlin, Germany\\
        E-mail: \email{ilgenfri@physik.hu-berlin.de}}

\author{H. Perlt\\
        Institut f\"ur Theoretische Physik, Universit\"at Leipzig, D-04009 Leipzig, Germany\\
        E-mail: \email{holger.perlt@itp.uni-leipzig.de}} 

\author{A. Schiller\\
        Institut f\"ur Theoretische Physik, Universit\"at Leipzig, D-04009 Leipzig, Germany\\
        E-mail: \email{arwed.schiller@itp.uni-leipzig.de}}

\abstract{By means of Numerical Stochastic Perturbation Theory (NSPT), we calculate the lattice 
gluon propagator up to three loops of perturbation theory in the limits of infinite volume and 
vanishing lattice spacing. Based on known anomalous dimensions and a parametrization of both 
the hypercubic symmetry group $H(4)$ and finite-size effects, we calculate the non-leading-log
and non-logarithmic contributions iteratively, starting with the first-loop expression.}

\FullConference{The XXVIII International Symposium on Lattice Field Theory, Lattice2010\\
		June 14-19, 2010\\
		Villasimius, Italy}

\begin{document}

\section{Introduction}

This talk was given as a progress report summarizing our work on the 
perturbative gluon propagator that has, in the meantime, published  
in Ref.~\cite{DiRenzo:2010cs}. 
This work is a continuation of Ref.~\cite{DiRenzo:2009ni} 
that was focused on the ghost propagator. The Monte Carlo study of both 
propagators (for a recent discussion see~\cite{Bogolubsky:2009dc}), which 
are closely related to each other by Schwinger-Dyson equations (SDE)~\cite{Alkofer:2000wg},
has attracted much attention outside the lattice community by phenomenologists
working in infrared QCD and hadron physics. For many reasons, the (covariant) 
Landau gauge propagators have been the centre of interest. The importance
for the confinement problem has been discussed in Ref.~\cite{Alkofer:2008bs}.

Taken together, both propagators provide us with a definition
and the momentum dependence of the running coupling $\alpha_s(q^2)$ 
directly based on the ghost-gluon vertex. 
This has recently turned out to be an interesting frame to accurately measure 
$\alpha_s(M_z)$ from 
propagators~\cite{Sternbeck:2007br,vonSmekal:2009ae,Sternbeck:2010xu}.

A simple connection between the two propagators exists in the extreme infrared, 
powerlike in a scaling and massive in a decoupling version. Only the latter is 
found on the lattice, while both solutions can be accommodated in the SDE approach. 
Some ideas exist today~\cite{Maas:2009se} about how to handle the 
Gribov ambiguity such that the 
scaling solution could eventually be reproduced on the lattice.

In a wider sense, the effect of nontrivial vacuum structure (vortices,
instantons) is manifest~\cite{Langfeld:2001cz,Boucaud:2003xi} 
also in the gluon propagator,
in the intermediate momentum range around ${\cal O}(1 \mathrm{~GeV})$
where the SDE approach suffers from truncation ambiguities and where 
nonperturbative lattice calculations are unrivalled. In order to understand 
the onset of nonperturbative effects, it is important to approach this momentum 
range from high momenta within higher-order perturbation theory. Some of us 
have started such a program a couple of years ago~\cite{Ilgenfritz:2007qj}. 
While ordinary diagrammatic lattice perturbation theory soon gets too involved 
to be pursued, Numerical Stochastic Perturbation Theory 
(NSPT, for a recent review see Ref.~\cite{DiRenzo:2004ge} and references therein), 
provides a powerful tool to perform high-loop computations. 
One has to run coupled Langevin simulations on the lattice and to perform the 
necessary limits: Langevin time step $\epsilon \to 0$, volume $V \to \infty$ and 
lattice spacing $a \to 0$.

The infinite-volume and continuum extrapolation part of the program has 
been satisfactorily achieved in Refs.~\cite{DiRenzo:2009ni,DiRenzo:2010cs},
such that the otherwise difficult to access non-logarithmic contributions 
became calculable. We refer the interested reader to these papers for more 
details about the technique of NSPT and the procedure to take the needed 
limits. The method to perform the $V \to \infty$ and $a \to 0$ limits
simultaneously has been outlined for the first time in Ref.~\cite{DiRenzo:2009ni} 
for the ghost propagator and applied to the gluon propagator in 
Ref.~\cite{DiRenzo:2010cs}. There we have attempted to compare the perturbative
results summed to the presently known order with the results of Monte Carlo
simulations. In this short account of our work we concentrate on the method
to extract all non-leading-log and non-logarithmic coefficients of the gluon
dressing function from NSPT data.


\section{The gluon dressing function}

Recalling that the gluon propagator has to be color-diagonal and symmetric 
in the Lorentz indices, its tensor structure in the continuum can be written as

\vspace*{-0.1cm}
\begin{equation}
D_{\mu\nu}^{ab}(p) = \delta^{ab}\Bigg[\Bigg(\delta_{\mu\nu}-\frac{p_{\mu}p_{\nu}}{p^2}\Bigg)
D(p^2)+\frac{p_{\mu}p_{\nu}}{p^2}\frac{F(p^2)}{p^2}\Bigg] \; ,
\end{equation} 
where the transverse part $D(p^2)$ and the longitudinal part $F(p^2)$ have been 
introduced. The latter vanishes in the Landau gauge.
The lattice counterpart $D_{\mu\nu}^{ab}(p(k))$$~\!$\footnote{$~\!$$p(k)$ 
components are defined 
as $p_{\mu}(k_{\mu})=2\pi k_{\mu}/L_{\mu}$, with $L_{\mu}=aN_{\mu}$ the lattice 
size along direction $\mu$ 
and $k_{\mu}\in[0,N_{\mu}-1]$. In what follows, the subscript $\mu$ on $N$ and 
$L$ will be dropped 
since simulations were performed on symmetric lattices.} 
of Eq.~(2.1) generally contains more terms 
on the r.h.s. due to the lower symmetry. The quantity under study, surviving the 
continuum limit, is 

\vspace*{-0.1cm}
\begin{equation}
D(p(k)) = \frac{1}{3}\sum_{\mu=1}^4D_{\mu\mu}(p(k)) \; ,
\end{equation}  
where color indices are dropped from now on to ease the notation. 
Eq.~(2.2) obviously corresponds to the transverse propagator.
In the following we use the $n^{th}$-order dressing function $J_{(n)}(p)$ 
defined as

\vspace*{-0.1cm}
\begin{equation}
J_{(n)}(p) = p^2D_{(n)}(p(k)) \; .  
\end{equation}


\section{The non-logarithmic contributions to Z}

In the RI'-MOM scheme, the relation between the bare dressing function 
$J_0(p,a,\alpha_{RI'})$ and its 
renormalized counterpart $J_{RI'}(p,\mu,\alpha_{RI'})$ is given 
by$~\!$\footnote{$~\!$Formulae in this section 
closely mimic those in section 4 of \cite{DiRenzo:2009ni} which we suggest 
the interested reader to consult.}

\vspace*{-0.1cm}
\begin{equation}
J_0(p,a,\alpha_{RI'})\ \! =\ \!Z(a,\mu,\alpha_{RI'})\ \!J_{RI'}(p,\mu,\alpha_{RI'}) \; ,
\end{equation}
with the renormalization condition 
$\left.J_{RI'}(p,\mu,\alpha_{RI'})\right|_{p^2=\mu^2}=1$ 
and with $\alpha_{RI'}=g^2(a\mu)/16\pi^2$. More in detail, the objects 
in Eq.~(3.1) can be perturbatively expanded as

\vspace*{-0.1cm}
\begin{eqnarray}
Z(a,\mu,\alpha_{RI'}) &=& 1 + \sum_{n>0}\alpha_{RI'}^n\sum_{j=0}^n z_{\ \!n,j}^{RI'}\ \!\log^j(a\mu)\ , \\
J_0(p,a,\alpha_{RI'}) &=& 1 + \sum_{n>0}\alpha_{RI'}^n\sum_{j=0}^n z_{\ \!n,j}^{RI'}\ 
\!\bigg(\frac{1}{2}\log(ap)^2\bigg)^{\!\!j}\ , \\
J_{RI'}(p,\mu,\alpha_{RI'}) &=& 1 + \sum_{n>0}\alpha_{RI'}^n\sum_{j=0}^n z_{\ \!n,j}^{RI'}\ 
\!\bigg(\frac{1}{2}\log(p\mu^{-1})^2\bigg)^{\!\!j} \; .
\end{eqnarray}
By plugging the perturbative expansions above into Eq.~(3.1) and by 
converting the renormalized 
coupling into the bare one, $\alpha_0=N_c/(8\pi^2\beta)$, the bare dressing 
function can be written as

\vspace*{-0.1cm}
\begin{equation}
J_0(p,a,\beta) = 1 + \sum_{n>0}\frac{1}{\beta^n}\sum_{j=0}^n J_{\ \!n,j}\ \!\log^j(ap)^2 \; ,
\end{equation}
where the coefficients $J_{\ \!n,j}$ are related to the $z_{\ \!n,j}^{RI'}$'s 
in Eq.~(3.2). 
As far as the logarithmic contributions are concerned, their coefficients 
depend on the anomalous 
dimension of the gluon field and the $\beta-$function 
(see \cite{Gracey:2003yr}).
The purpose of this work is to compute the $J_{\ \!n,0}$'s which are related to 
the  $z_{n,0}^{RI'}$'s by

\vspace*{-0.1cm}
\begin{eqnarray}
J_{1,0} &=& 0.03799544\ \!z_{1,0}^{RI'}\ ,\\
J_{2,0} &=& 0.10673710\ \!z_{1,0}^{RI'} + 0.00144365\ \!z_{2,0}^{RI'}\ ,\\
J_{3,0} &=& 0.375990\ \!z_{1,0}^{RI'} + 0.00811105\ \!z_{2,0}^{RI'} + 0.0000548523\ \!z_{3,0}^{RI'} \; .
\end{eqnarray} 


\section{The fitting procedure}

We can isolate the $n^{th}-$loop contribution in Eq.~(3.5) 
(without power of $\beta^{-n}$) and write

\vspace*{-0.1cm}
\begin{equation}
J_{(n)}(p,a) = J_{n,0}(ap) + \sum_{j=1}^n\ \!J_{n,j}\ \!\log^j(ap)^2 \; ,
\end{equation}
where, by recalling the existence of irrelevant lattice 
artifacts, $J_{n,0}(ap)$ can be decomposed as

\vspace*{-0.25cm}
\begin{equation}
J_{n,0}(ap) = J_{n,0} + c_{n,1}(ap)^2 + c_{n,2}\frac{(ap)^4}{(ap)^2} + c_{n,3}(ap)^4 + \ldots \ \ \ \ \ \ \ 
[(ap)^m\equiv\sum_{\mu}(ap_{\mu})^m] \; , 
\end{equation}
where $J_{n,0}$ is the $n^{th}-$loop constant we want to compute. 
Taking into account also finite-size effects, 
$J_{n,0}(ap)$ has to be replaced by $J_{n,0}(ap,pL)$ with

\vspace*{-0.2cm}
\begin{eqnarray}
J_{n,0}(ap,pL) &=& J_{n,0}(ap) + [J_{n,0}(ap,pL) - J_{n,0}(ap)] \equiv J_{n,0}(ap) + \delta J_{n,0}(ap,pL) = \nonumber \\
&=& J_{n,0}(ap) + \delta J_{n,0}(0,pL) \equiv J_{n,0}(ap) + \delta J_{n,0}(pL) \ , 
\end{eqnarray}
where, in the last two steps, $ap$ corrections on $pL$ effects are assumed to 
be \emph{corrections on corrections} 
and have been neglected. Furthermore we assume that the point $p_{\max}$ 
corresponding to the largest $(ap)^2$ 
on the largest lattice size  $L_{\max}$ is such that 
$\delta J_{n,0}(p_{\max}L_{\max})=0$.

When treating $pL$ contributions, it is useful to notice that, 
since $p_{\mu}L = p_{\mu}aN = 2\pi k_{\mu}$, 
any given 4-tuple of integers $(k_1,k_2,k_3,k_4)$ has the same finite-size 
corrections to $J_{n,0}(ap)$ 
on different lattice sizes. Therefore, the steps in data-analysis at any 
loop order $n$ can be summarized as follows: 

\vspace*{0.3cm}

\begin{itemize}

\item a suitable window $[(ap)^2_{\min},(ap)^2_{\max}]$ is identified 
where a sufficient number of data points 
is available on different lattice sizes;

\item the logarithmic contributions are subtracted from the bare dressing 
function in order to get $J_{n,0}(ap,pL)$;

\item $J_{n,0}(ap,pL)$ is fitted according to the procedure sketched above: 
from Eq.~(4.2) we compute the desired $J_{n,0}$. 

\end{itemize}

\vspace*{0.3cm}

\noindent Obviously, only results stemming from stable fits have to be taken into account.


\section{Results}

Figs. 1, 3 and 5 show an example of the fitting procedure outlined above: 
\begin{figure}[!ht]
 \hfill
 \begin{minipage}[t]{.47\textwidth}
 \begin{center}
   \hspace*{-0.5cm}%
   \includegraphics[height=5.0cm]{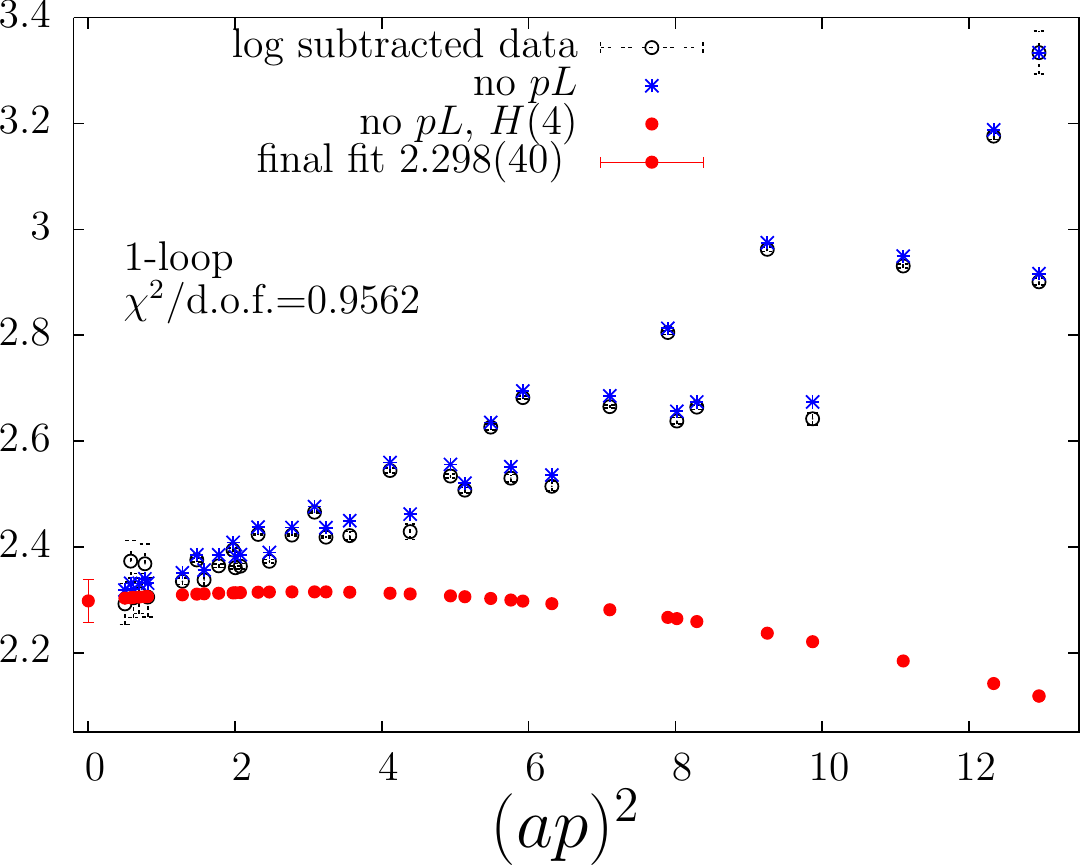}
   \caption{1-loop data vs. $(ap)^2$: the different sets of points stem from the various steps of the analysis (see section 4).}
   \label{Fig.1}
  \end{center}
 \end{minipage}
 \hfill
 \begin{minipage}[t]{.47\textwidth}
  \begin{center}
  \vspace*{-5.1cm}
   \includegraphics[height=5.1cm]{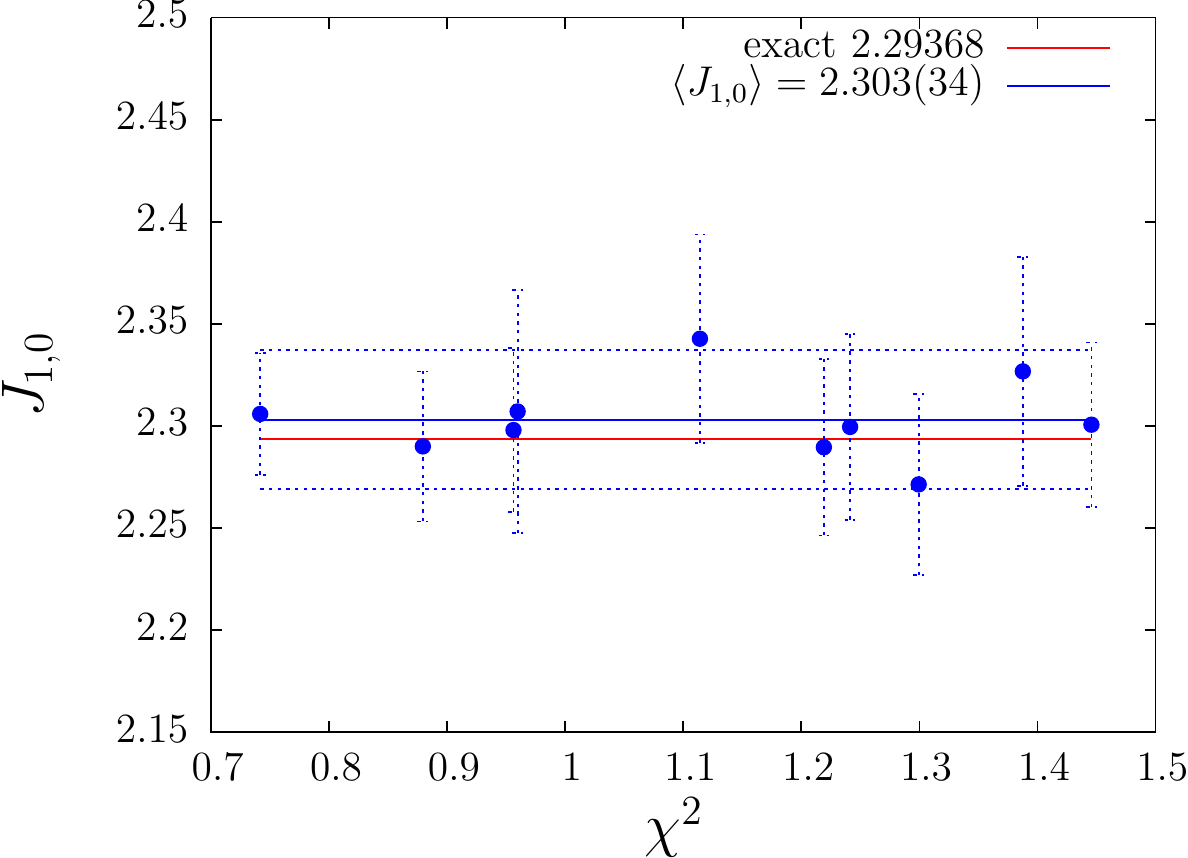}
   \caption[a]{$J_{1,0}$ results vs. $\chi^2$: the dotted blue line is the average of the 10 blue points,
               the continuous red one is the analytical result.}
   \label{Fig.2}
  \end{center}
 \end{minipage}
 \hfill
\end{figure}
\begin{figure}[!htb]
 \hfill
 \begin{minipage}[t]{.47\textwidth}
 \begin{center}
   \hspace*{-0.5cm}%
   \includegraphics[height=5.0cm]{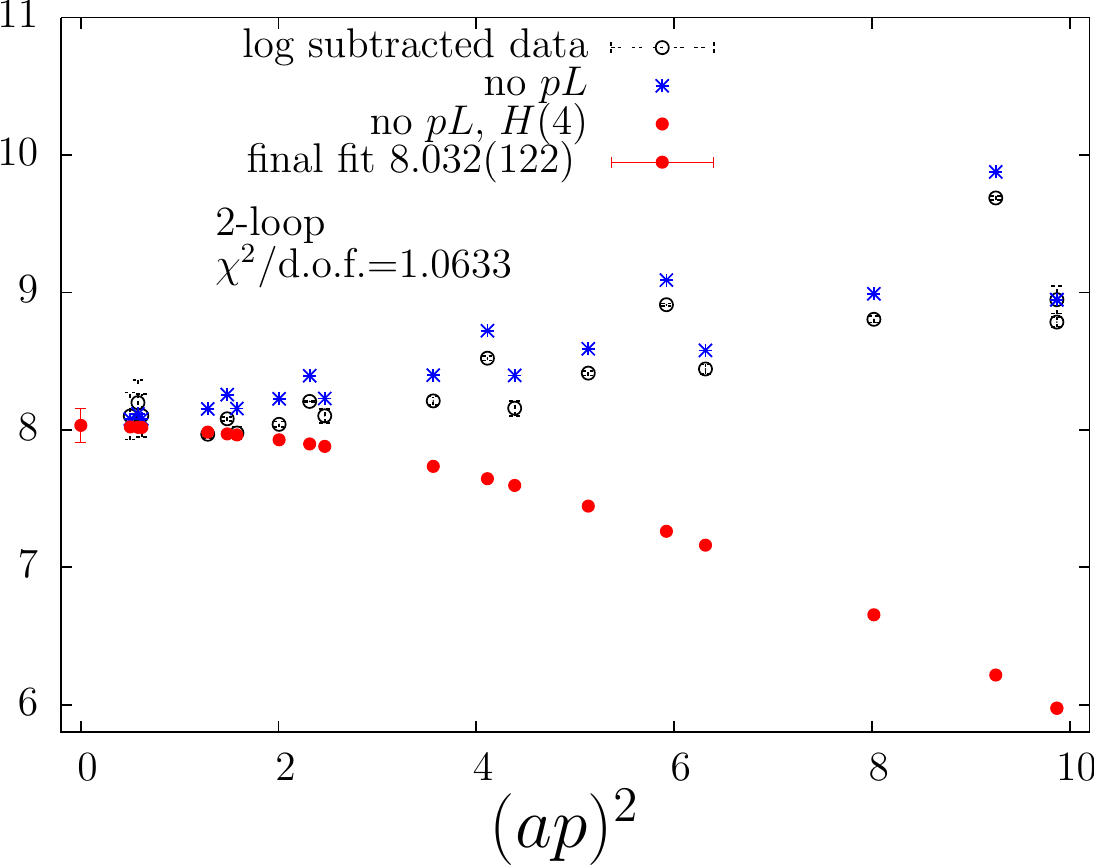}
   \vspace*{-0.5mm}
   \caption{Same as in Fig. 1 for 2-loop data.}
   \label{Fig.3}
  \end{center}
 \end{minipage}
 \hfill
 \begin{minipage}[t]{.47\textwidth}
  \begin{center}
  \vspace*{-5.1cm}
   \includegraphics[height=5.1cm]{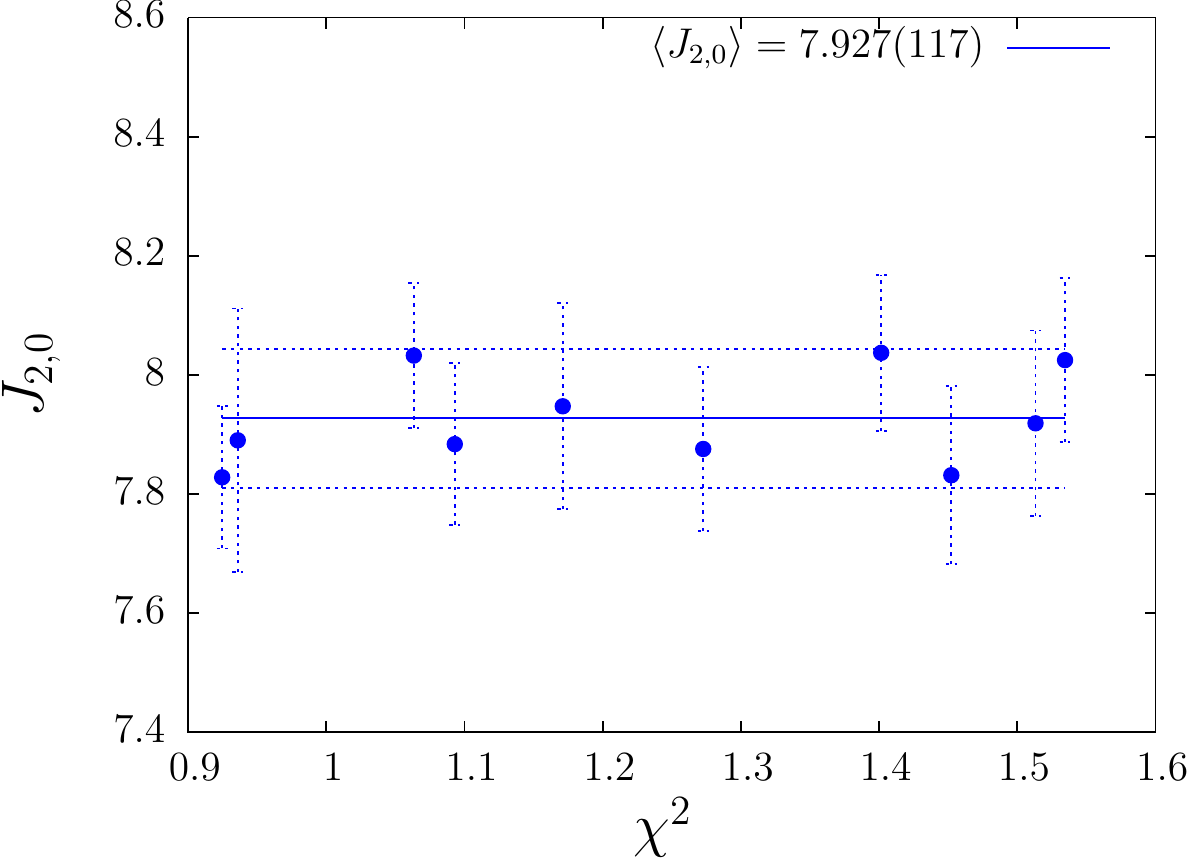}
   \caption[a]{$J_{2,0}$ results vs. $\chi^2$: see caption of Fig. 2.}
   \label{Fig.4}
  \end{center}
 \end{minipage}
 \hfill
\end{figure}
the black dots stand for log-subtracted data, 
blue stars for data after finite-size effects have been removed while red 
spots for data after further removal of hypercubic effects.   


\vspace*{0.35cm}
\begin{table}[h]
\begin{center}
\begin{tabular}{|c|c|c|c|}
\hline
Loop order n &  1  &  2  &  3  \\ 
\hline
$J_{n,0}$  &  2.30(3) & 7.92(12) & 31.7(5)\\
\hline
\end{tabular}
\vspace*{0.1cm}
\caption{Results for the constant contributions to $Z$ up to the third loop. 
Compare the 1-loop {\it analytical} result $J_{1,0}=2.29368$.}
\label{Table 1}
\end{center}
\end{table}


\vspace*{-0.3cm}
In general, there could be various fitting functions that fulfill the stability 
requirement mentioned at the end of section 4: the results we quote come from 
averaging the values obtained from the fits with the 10 lowest $\chi^2$ 
at every perturbative order: Figs. 2, 4 and 6 contain these values and their 
average with their errorbars. 
The final outcome of the analysis sketched in the above is contained in Table 1.

Resumming the perturbative series for $J_0(a,p,\beta)$ up to the third loop, 
we get

\vspace*{-0.02cm}
\begin{eqnarray}                                                                                                                                        J^{\rm 3-loop}(p,a, \beta) &=& 1 + \frac{1}{\beta}\,\Bigl( -0.24697\, \log (ap)^2 + 2.29368 \Bigr) + \nonumber \\                                     &+& \frac{1}{\beta^2}\,\Bigl(0.08211\, \left( \log (ap)^2 \right)^2 - 1.48445\log (ap)^2 + 7.93(12) \Bigr) + \\                          
&+& \frac{1}{\beta^3}\,\Bigl( -0.02964\, \left( \log (ap)^2 \right)^3 + 0.81689 \,\left( \log (ap)^2 \right)^2 + \nonumber\\ 
&-& 8.13(3) \, \log (ap)^2 + 31.7(5) \Bigr) \; , \nonumber                                                                                              
\end{eqnarray}
\vspace*{-0.1cm}

which can then be converted to the RI'-MOM scheme by using the formulae of section 3.


\begin{figure}[t]
 \vspace*{0.5cm}
 \hfill
 \begin{minipage}[t]{.47\textwidth}
 \begin{center}
   \hspace*{-0.5cm}%
   \includegraphics[height=5.0cm]{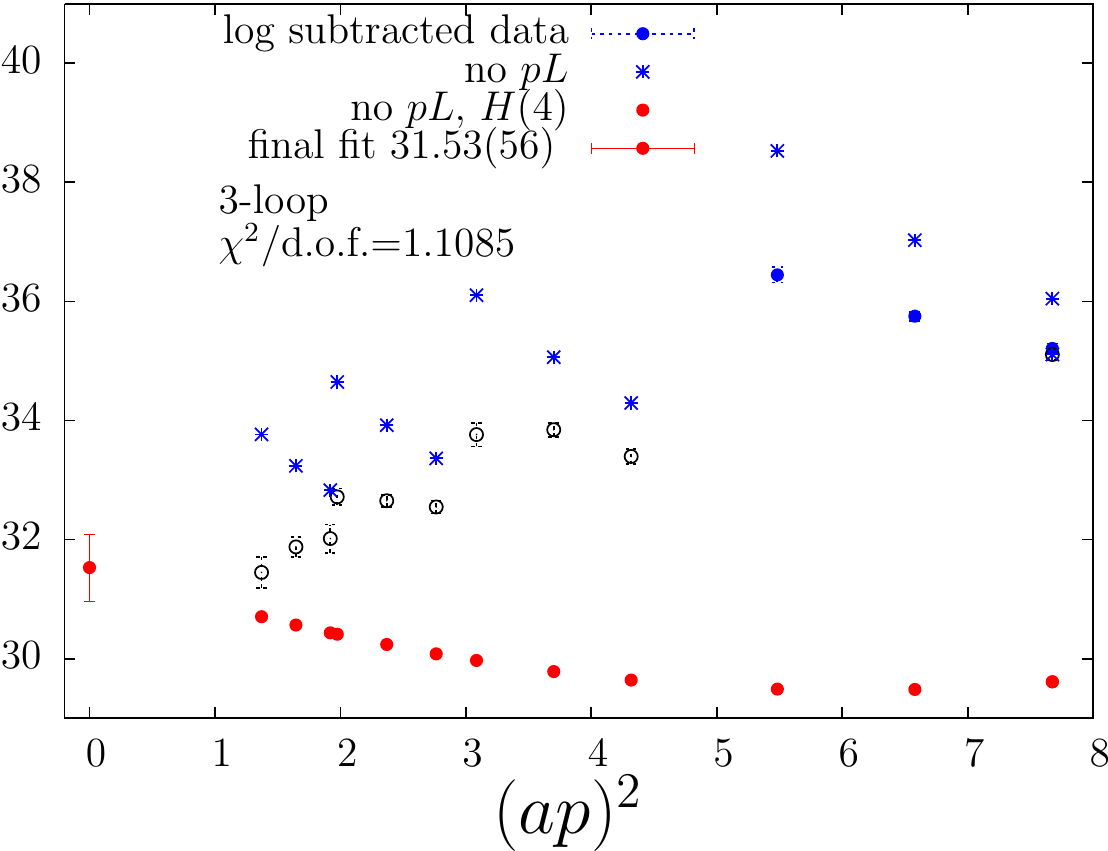}
   \vspace*{-0.5mm}
   \caption{Same as in Fig. 1 for 3-loop data.}
   \label{Fig.5}
  \end{center}
 \end{minipage}
 \hfill
 \begin{minipage}[t]{.47\textwidth}
  \begin{center}
  \vspace*{-5.1cm}
   \includegraphics[height=5.1cm]{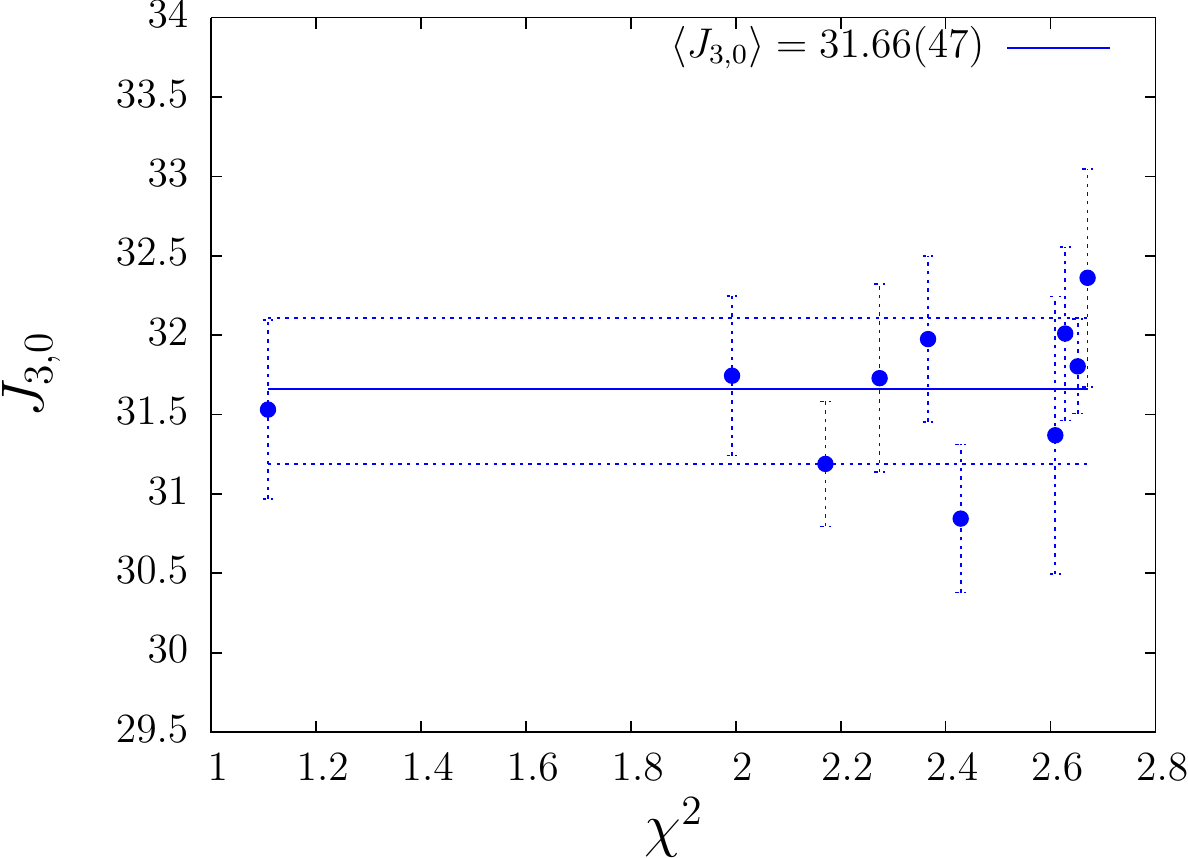}
   \caption{$J_{3,0}$ results vs. $\chi^2$: see caption of Fig. 2.}
   \label{Fig.6}
  \end{center}
 \end{minipage}
 \hfill
\end{figure}


\section{Conclusion}

In this work we presented an algorithm to compute the non-logarithmic parts of 
infrared divergent quantities 
in NSPT in the infinite-volume limit using the example of the gluon propagator. 
The one-loop result coincides 
with the well-known result from the diagrammatic approach~\cite{Kawai:1980ja}.
The corresponding two- and three-loop finite constants have been computed for 
the first time.

We can use the  dressing function obtained with NSPT at finite lattice sizes 
to investigate the perturbative background of the corresponding Monte-Carlo 
calculated propagators. This has been done in some detail in~\cite{DiRenzo:2010cs}.


\section*{Acknowledgements}

This work has been supported by DFG under contract SCHI 422/8-1, DFG SFB/TR 55,
by I.N.F.N. under the research project MI11 and by the Research Executive Agency 
(REA) of the European Union under Grant Agreement number PITN-GA-2009-238353 
(ITN STRONGnet).


\end{document}